\documentclass[12pt]{article}
\usepackage{latexsym}
\usepackage{amsmath}
\usepackage{indentfirst}
\usepackage{cite} % Orders and compresses citation strings
\long\def\ca#1\cb{} %Use for commenting out: \ca...\cb

\newcommand{\AND}{\mbox{\small AND}}

\newcommand{\becs}{\begin{cases}}
\newcommand{\bem}{\begin{matrix}}

\newcommand{\dyad}[2]{|#1\rangle\langle#2|}

\newcommand{\encs}{\end{cases}}
\newcommand{\enm}{\end{matrix}}

\newcommand{\ket}[1]{|#1\rangle }

\newcommand{\lra}{\leftrightarrow }

\def\outl#1{\par{\medskip\noindent\hspace*{0.1cm}\bf
      \mathversion{bold}#1\mathversion{normal}\smallskip} }
   \def\xa{} \def\xb{}  

 \def\outl#1{}\def\xa{}\def\xb{}
\ca
 \def\outl#1{\par{\medskip\noindent\hspace*{.5cm}\bf
      \mathversion{bold}#1\mathversion{normal}\smallskip} }
 \long\def\xa#1\xb{} %This comments out segments from \xa to the next \xb
  
\cb
\numberwithin{equation}{section}

\begin{document}

\title{Comment on Chiribella et al., Phys. Rev. Lett. 132 (2024) 190201
 [arXiv:2301.10885]}

\author{Robert B. Griffiths\thanks{Electronic address: rgrif@cmu.edu}\\
Department of Physics\\
Carnegie Mellon University\\
Pittsburgh, PA 15213}

\date{Version of 29 May 2024}
\maketitle

\begin{abstract}
  The article ``Bell Nonlocality in Classical Systems Coexisting with Other
  System Types'' by Chiribella et al. defines ``classical'' in a
  quantum context in a way that ignores noncommuting quantum projectors, and is
  hence inconsistent with Hilbert-space quantum theory.
\end{abstract}

\xb%
\outl{CLASSICAL is central concept in \cite{ChGS24}. Noncommutation separates
HSQM from classical. PDI. Contrast $AB=BA$ vs $AB \neq AB$ }%
\xa%

A central concept in Chiribella et al.\ \cite{ChGS24} is \emph{classical} as used in a quantum
context, the simplest situation being a few qubits. A key feature that
distinguishes Hilbert-space mathematics from that used in classical physics is
the noncommutation of quantum operators, in particular the projectors that
define quantum properties, such as $S_x=+\hbar/2$ for a spin-half particle; see
Sec.~III.5 of \cite{vNmn55}. A collection of mutually-orthogonal projectors
which sum to the identity operator, a \emph{projective decomposition of the
  identity} (PDI), is the quantum analog of a classical phase space, allowing
properties expressed by these projectors or their sums to be combined in
classical fashion, e.g., the operator $AB=BA$ in the commuting case can be
interpreted as ``$A\, \AND\, B$'', unlike the case $AB \neq BA$, for which
there is no classical analog.

\xb%
\outl{ `measurement $\lra$ pre-existing property' in
\cite{ChGS24} not satisfactory for defining `classical' in Qm context}%
\xa%

\xb%
\outl{Bell ``Against Measurement'': Textbook approach does not work}%
\xa%

By contrast, \cite{ChGS24} starts with the notion that ``classical measurements
reveal pre-existing properties of the measured system'', and then assumes that
this can be used to somehow define ``classical'' even in a \emph{quantum}
context, for a measurement carried out on a microscopic quantum system. This is
unsatisfactory for several reasons; Among others it runs straight into the
infamous \emph{measurement problems} of quantum foundations. Such as the fact
that textbook quantum measurements do not actually measure anything, as pointed
out by Bell \cite{Bll901} a quarter century ago. Current textbooks are no
better.

\xb%
\outl{Need Qm description with both earlier property + measurement outcome}%
\xa%

\xb%
\outl{CH provides this. No CH references in \cite{ChGS24} bibliography}%
\xa%

What is needed is a consistent way of viewing a quantum measurement such that
both the earlier state of the measured system as well as the later measurement
outcome can be described in quantum-mechanical terms using appropriate
projectors at the different times, leading to a one-to-one correlation between
the prior property and the later outcome.
Such a description is possible in the case of projective measurements using
quantum histories and an analysis based on the consistent/decoherent histories
(CH) formulation, see \cite{Grff17b}. I am not aware of any other approach to
quantum foundations that achieves this, and am very disappointed that not a
single reference to any CH publication is included among the 60 items making up
the bibliography of \cite{ChGS24}.

\xb%
\outl{CH prohibits combining noncommuting projectors = SFR}%
\xa%

\xb%
\outl{Quasiclassical frameworks and pointer measurement problem}%
\xa%

Among other things, CH absolutely prohibits using noncommuting projectors, be
it at a single time or in quantum histories, in a single quantum description,
such as a single run of an experiment. This \emph{single framework rule} is the
central feature which allows CH to resolve numerous quantum paradoxes,
including the well-known measurement paradox that arises in trying to reconcile
unitary time development with a definite location of an apparatus output
pointer. The CH approach uses a quasiclassical framework \cite{GMHr93} for the
measurement outcome, one in which projectors on macroscopic quantum properties
(e.g., pointer positions) commute for all practical purposes. The connection of
these outcomes with microscopic properties is straightforward in the case of
projective measurements \cite{Grff17b}, whereas similar, but more
complicated, conclusions apply in the case of POVMs.

\xb%
\outl{Qm analysis of Ats' ``classical'' in case of 2 entangled qubits}%
\xa%

\xb%
\outl{CH: Projective measurements reveal prior micro properties using PDI at
  time \emph{later} than initial state $\ket{\Phi}$, but PDI projectors
  incompatible with $[\Phi]$ }%
\xa%

\xb%
\outl{This analysis used in lab, but absent from textbooks}%
\xa%

Let us nonetheless see if the authors' concept of ``classical'' can be analyzed
in proper quantum-mechanical terms, starting with the case of just two qubits
in a fully entangled Bell state, e.g., one of the pairs connected by a wiggly
line in their Fig.~2. In the case of projective measurements there is a very
general argument, see \cite{Grff17b}, leading to the result that each
macroscopic outcome is uniquely correlated with a microscopic property
represented by a suitable quantum projector at an earlier time, provided one
employs an appropriate framework. These microscopic projectors form a PDI, and
the associated microstates correspond with the later outcomes of projective
measurements, in agreement with the authors' notion of classical realism.
Furthermore this is so when both Alice and Bob simultaneously measure the two
qubits in an entangled state. Note, however, that the properties revealed by
the later measurements are represented by projectors at an earlier time
\emph{preceding} the measurements, but \emph{later} than that at which the
initial entangled state $\ket{\Phi}$ was prepared. And the projectors for the
measured properties do not commute with the projector $\dyad{\Phi}{\Phi}$
representing the initial entangled state. 
Such a notion of properties later revealed by measurements at an intermediate
time between preparation and measurement is absent (as Bell complained) from
typical textbook treatments of measurements. In the physics laboratory
experimenters routinely understand a (macroscopic) measurement outcome as
\emph{caused} by an earlier microscopic property which their apparatus,
properly tested and calibrated, has revealed. For a particular and quite simple
illustration see Sec.~III of \cite{Grff24}.

\xb%
\outl{Argument in \cite{ChGS24}: violation of CHSH for system in its Fig.~2}%
\xa%

\xb%
\outl{Bell's use of classical HVs contrary to vN. Expt showed vN was right.
\cite{Grff20} has details}%
\xa%

The authors' argument in the case of the system in their Fig.~2 proceeds by
exhibiting a violation of the CHSH inequality \cite{CHSH69}, a particular form
of Bell's paradox, in the case of an overall system in which there are two
2-qubit entangled pairs. with one element of each pair in Alice's domain and
the other in Bob's. Since the usual CHSH inequality is violated in experimental
tests using just one such pair, it is hardly surprising that it fails for two
pairs. One way to view the source of such violations is Bell's improper use of
\emph{classical} hidden variables in a situation where their quantum
counterparts are noncommuting projectors, something von Neumann regarded as
contrary to quantum principles. As is well know, experiment has shown that von
Neumann was right and Bell was wrong. Detailed analyses of the situation will
be found in \cite{Grff20}, among other references that pay serious attention to
noncommutation.

\xb%
\outl{SUMM: \cite{ChGS24} does not take noncommutation into account}%
\xa%

\xb%
\outl{CH may not be unique or best way for consistent QM, but it resolves many
  paradoxes and deserves attention}%
\xa%

In summary, the basic problem with \cite{ChGS24} is the failure to understand
that noncommutation of projectors is at the heart of the difference between
classical and quantum mechanics, and must be taken into account in order to
make sense of the latter The CH approach may not be the only way to carry out a
logically consistent quantum analysis, but it has the advantage that it has
been around for some time, and has been shown to resolve many quantum paradoxes
which other quantum interpretations are unable to handle, as in Chs.~20 to 25
of \cite{Grff02c}. For a brief introduction to CH see \cite{Grff19b}. It may
well be possible to improve on it, but this would best be done by first paying
some attention to what has already been accomplished.

% Acknowledgements
\xb
\section*{Acknowledgements}
\xa

The author is grateful to
Carnegie-Mellon University and its Physics Department for continuing support of
his activities as an emeritus faculty member.

% The following two commands for using .bib file
% Comment them out, insert name.bbl file, to make latex file self-contained
%\bibliographystyle{unsrt}
%\bibliography{/home/rgrif/qms/bibs/main,notes}
%\bibliography{/home/rgrif/qms/bibs/main}

\end{document}